\documentclass[]{spie}  

\usepackage{graphicx}

\begin{document}

\thispagestyle{empty}
\renewcommand{\thefootnote}{\fnsymbol{footnote}}

\begin{flushright}
{\small
SLAC--PUB--12509\\
May, 2007\\}
\end{flushright}

\vspace{.8cm}

\begin{center}
{\bf\large   
Studies of Cosmic Rays with GeV Gamma Rays
\footnote{This work was supported in part by the U.S. Department of Energy under Grant DE-AC02-76SF00515.}}

\vspace{1cm}

Hiroyasu~Tajima,$^{1}$ Tuneyoshi~Kamae,$^{1}$ Stefano~Finazzi,$^{2}$ Johann~Cohen-Tanugi$^{1}$ and James~Chiang$^{1,3}$
\medskip

(1) Stanford Linear Accelerator Center,
  2575 Sand Hill Road, Menlo Park, CA 94025, USA\\
(2) Scuola Normale Superiore, I-56100 Pisa, Italy\\
(3) Center for Research and Exploration in Space Science and Technology (CRESST), \\University of Maryland, Baltimore County, 1000 Hilltop Circle, Baltimore, MD 21250, USA

\end{center}

\vfill

\begin{center}
{\bf\large   
Abstract }
\end{center}

\begin{quote}
We describe the role of GeV gamma-ray observations with GLAST-LAT (Gamma-ray Large Area Space Telescope - Large Area Telescope) in identifying interaction sites of cosmic-ray proton (or hadrons) with interstellar medium (ISM).
We expect to detect gamma rays from neutral pion decays in high-density ISM regions in the Galaxy, Large Magellanic Cloud, and other satellite galaxies.
These gamma-ray sources have been detected already with EGRET (Energetic Gamma Ray Experiment Telescope) as extended sources (eg. LMC and Orion clouds) and GLAST-LAT will detect many more with a higher spatial resolution and in a wider spectral range.

We have developed a novel image restoration technique based on the Richardson-Lucy algorithm optimized for GLAST-LAT observation of extended sources.
Our algorithm calculates PSF (point spread function) for each event. 
This step is very important for GLAST-LAT and EGRET image analysis since PSF varies more than one order of magnitude from one gamma ray to another depending on its energy as well as its impact point and angle in the instrument. 
The GLAST-LAT and EGRET image analysis has to cope with Poisson fluctuation due to low number of detected photons for most sources. 
Our technique incorporates wavelet filtering to minimize effects due to the fluctuation.
Preliminary studies on some EGRET sources are presented, which shows potential of this novel 
image restoration technique for the identification and characterisation of extended gamma-ray sources.
\end{quote}

\bigskip


\vfill

\begin{center} 
{\it Invited Talk at} 
{\it International Workshop on ``Cosmic-Rays and High Energy Universe,'' \\
Aoyama-Gakuin University, Shibuya, Tokyo, Japan, March 5--6, 2007} \\



\end{center}

\clearpage
\pagestyle{plain}

\section{Introduction}
The origin of the cosmic rays has been a great mystery since their discovery by Victor Hess in 1912.  
Supernova remnants (SNRs) are considered to be the best candidates for the proton (hadron) acceleration up to 
the so-called ``knee" ($3\times10^{15}$ eV) in the cosmic-ray spectrum. 
Supports for this hypothesis are mostly circumstantial and theoretical rather than observational.
For example, the kinetic energy released by supernova explosion has been known to be comparable with the total energy in the Galactic cosmic ray\cite{Ginzburg64}.
Simple diffusive shock acceleration modeling of a typical SNR has given the maximum energy of accelerated particles to reach the "knee"(see for example a review article\cite{Blandford87}).
Synchrotron emission in the X-ray band observed in SN1006 by ASCA\cite{Koyama95} was the first strong indirect evidence 
for the existence of $\sim$100 TeV electrons in an SNR. 
Observations of TeV gamma rays from another SNR, RX J1713.7-3946, by CANGAROO\cite{Enomoto02} and by H.E.S.S.\cite{Aharonian04} 
were the first confirmed direct evidence for particle acceleration up to 100 TeV in a SNR.
Furthermore, H.E.S.S. provided the first gamma-ray image of RX J1713.7-3946, which is qualitatively consistent with that obtained in the X-ray band. This morphorogical coincidence has been interpreted that the accelerated particles are interacting with ISM or cosmic microwave backgrounds (CMB) near or in the shock region. 
However, the TeV spectrum observed by H.E.S.S. alone does not favor decisively a dominant gamma-ray emission mechanism (Compton up-scattering or $\pi^0$-decays following proton-proton interactions).
Hence, the proof of hadronic acceleration in SNRs is not yet conclusive and wait for new data from GLAST-LAT in the GeV energy range.

There is no consensus for acceleration sites of cosmic rays above the ``knee" of the cosmic-ray spectrum. Several astronomical objects such as gamma-ray bursts (GRBs), active galaxies or merging galaxy clusters 
have been put forward as possible candidates.
Observation of UHECRs (Ultra High-Energy Cosmic Rays) and UHE-neutrinos associated with known astronomical sources 
will provide convincing evidence for UHECR acceleration in these sources.
Unfortunately the phase space for UHECRs to reach a detector is rather small because charged particles are deflected by the extragalactic and interstellar magnetic field below $10^{19}$~eV. It is also known that hadronic cosmic rays cannot travel more than $\sim$20~Mpc above $\sim10^{19.8}$~eV due to production of the $\Delta$ resonance by interacting with CMB (known as GZK suppression\cite{GZK}).
Neutrinos are free from both constraints and allow us to probe distant UHECR acceleration sites if there is enough target material at the sites.
However, neutrinos are difficult to detect due to their small cross section.
In the case of gamma-ray observations, multi-wavelength analysis is critical to resolve between the leptonic and hadronic model of gamma-ray emission. Neutrino detection from the same gamma-ray sources wil provide definitive evidence for acceleration of protons in such sources.

\section{Cosmic Rays from Supernova Remnants}
\begin{figure}[bth]
\centering
\begin{tabular}{ll}
(a) & (b) \\
\includegraphics[height=5.3cm]{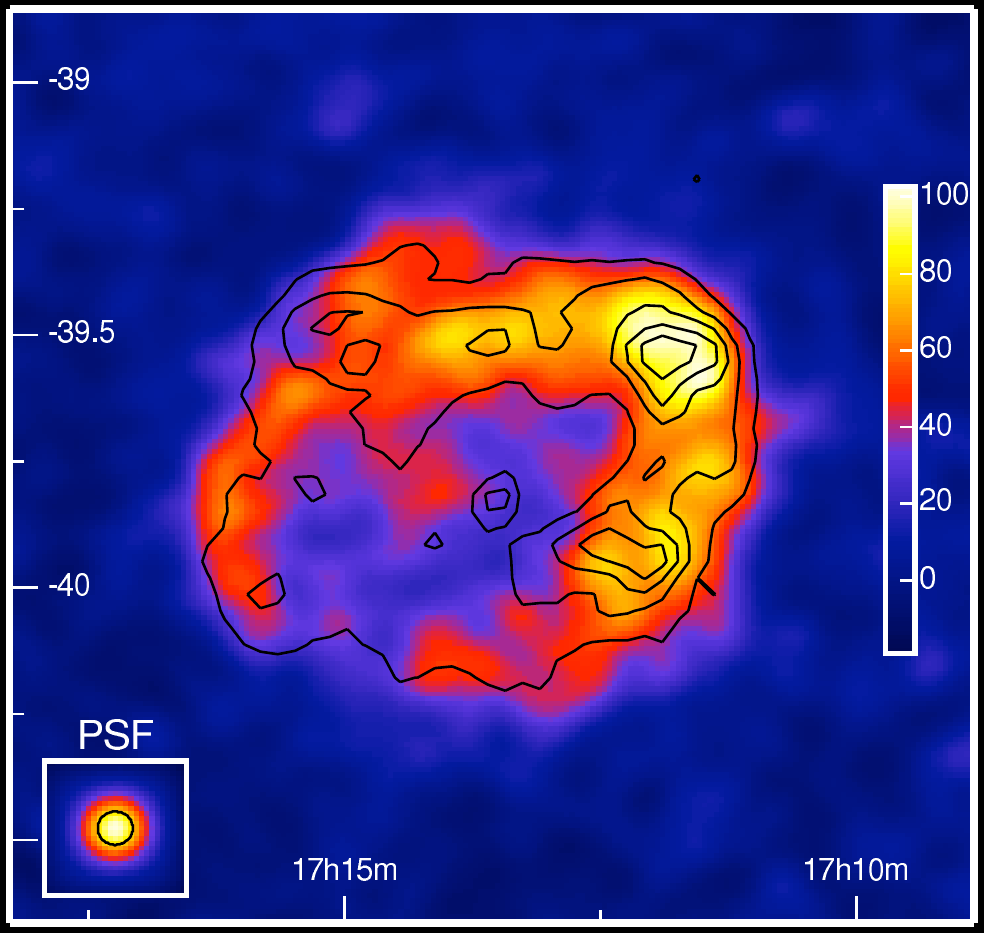} &
\includegraphics[height=5.3cm]{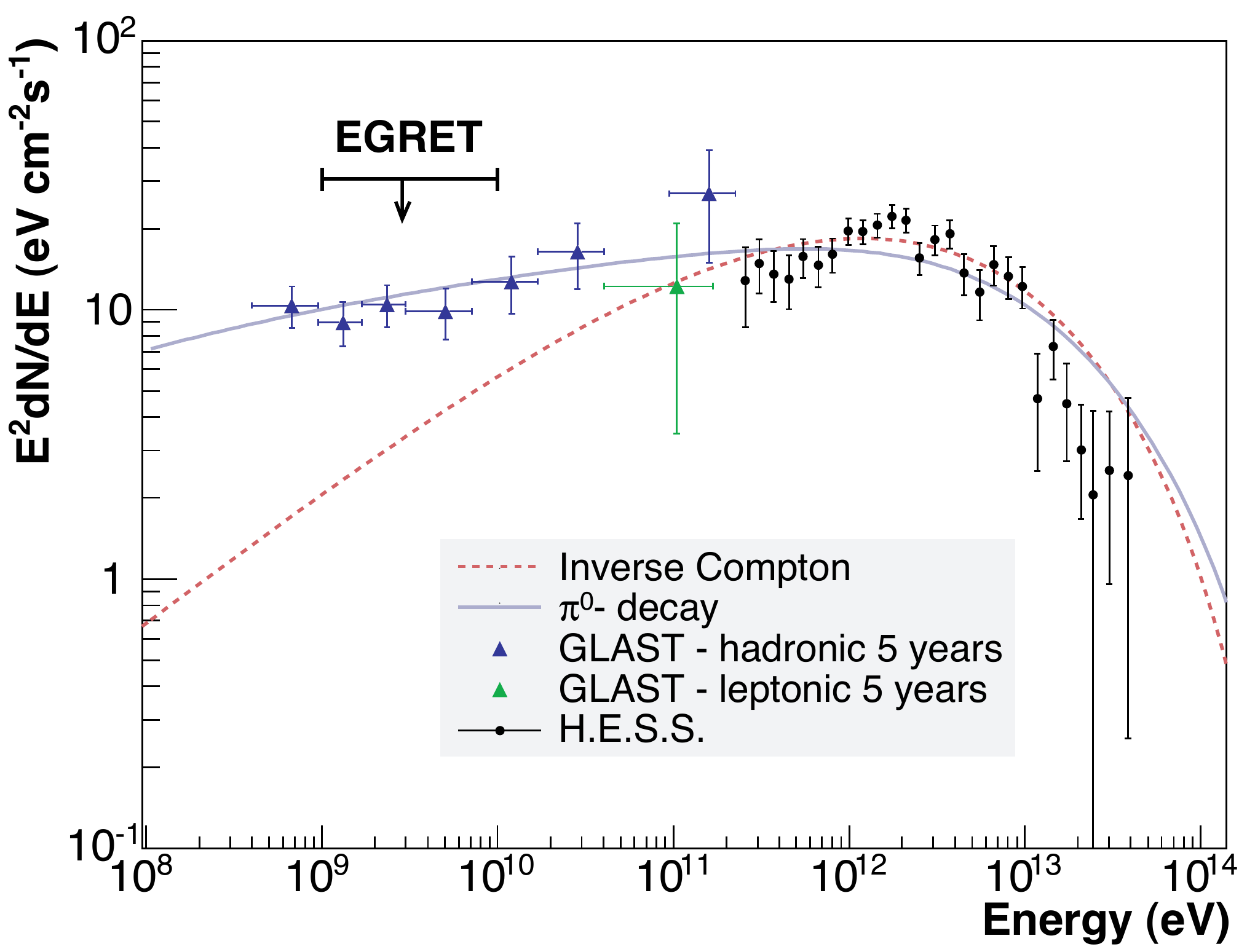}
\end{tabular}
\caption[]
{(a) Acceptance corrected gamma-ray excess image of the SNR, RX J1713.7-3946, observed by H.E.S.S.\cite{Aharonian07}. Contour of the ASCA X-ray (1--3~keV band) image\cite{Uchiyama02} is superimposed.
(b) Gamma-ray spectrum of RX J1713.7-3946 observed by H.E.S.S. Simulated GLAST-LAT measurements for hadronic and electronic models are also shown\cite{Funk}.}
\label{fig:HESS-RXJ1713}
\end{figure}

As described above, SNRs are the prime candidates for the origin of Galactic cosmic rays.
Recent observations of TeV gamma rays from RX J1713.7-3946 and RX J0852.0-4622, by H.E.S.S.\cite{Aharonian04,Aharonian06b} 
present undisputed evidence that electrons and/or protons are accelerated to at least $10^{14}$~eV, which is very close to the ``knee" energies.
Fig.~\ref{fig:HESS-RXJ1713}~(a) shows the latest TeV gamma-ray image of a shell-type SNR, RX J1713.7-3946, which is the first SNR imaged in TeV gamma rays.
Morphological similarity between X-ray\cite{Uchiyama02} and TeV gamma ray\cite{Aharonian07} as shown in Fig.~\ref{fig:HESS-RXJ1713}~(a) may suggest that Compton up-scattering of CMB photons to TeV by high energy electrons and positrons that produce the synchrotron radiation observed by ASCA.
The azimuthal profile of gamma-ray intensity observed by H.E.S.S. and CO intensity observed by NANTEN are not as well correlated as between H.E.S.S. and ASCA.
This may present some difficulties for the proton interaction model.
The spectrum of the H.E.S.S. observation does not necessarily agree with the prediction by the leptonic model\cite{Aharonian06}.
Furthermore, Hiraga {\it et al}.\cite{Hiraga05} pointed out that the radial profile of the X-ray mission cannot be explained by smooth spherical distributions expected from leptonic models.
In addition, they observed possible evidence of the positive correlation between the X-ray brightness and the absorption column density, which may indicate that molecular clouds are providing unaccelerated electrons (and protons) to the shock, resulting in higher electron population thus higher X-ray emissivity.
Using this scenario, the morphological similarity between TeV gamma ray and X-ray images can be explained as a similarity in proton and electron injection rate in the dense matter region, making a hadronic model a viable alternative to leptonic models.
More definitive evidence is required to conclusively determine the dominant gamma ray emission mechanism at RX J1713.7-3946.
Fig.~\ref{fig:HESS-RXJ1713}~(b) shows the H.E.S.S. measurement result of RX J1713.7-3946 along with simulated GLAST-LAT measurements for hadronic and leptonic cases\cite{Funk}.
It demonstrates that GeV gamma-ray observations by the GLAST-LAT can provide conclusive evidence on this matter since the spectra predicted by hadronic and leptonic models are clearly separated in the GLAST-LAT range ($\sim$100~MeV to 300~GeV).

\section{Cosmic-Ray Interactions with the Galactic Interstellar Medium}
Diffuse gamma-ray emissions from the Galactic plane provide rich information on Galactic cosmic rays, cosmic-ray propagation, Galactic interstellar medium (ISM), and cosmic-ray interactions with ISM and low-energy photons.
Hunter {\it et al.}\cite{Hunter97} fitted the EGRET gamma-ray spectrum in the Galactic center region with a model gamma-ray spectral shape obtained on a model proton-proton pion production cross-section and a model for the spatial 
distribution of pion gamma-rays projected to the Galactic coordinate.
The fit left the absolute normalization undetermined because matter density nor cosmic-ray intensity in the Galactic center region were not known.
The spectral shape of the model used in the publication reproduced the observed spectrum poorly when normalized at the peak in the $\nu F(\nu)$: the model left possible 50\% ``excess'' in the observed flux above 1~GeV.
This possible excess gamma-ray flux has led to many speculations including claims for dark matter detection.
\begin{figure}[bht]
\centering
\begin{tabular}{lll}
(a) & (b) \\
 & &
\includegraphics[height=2.4cm]{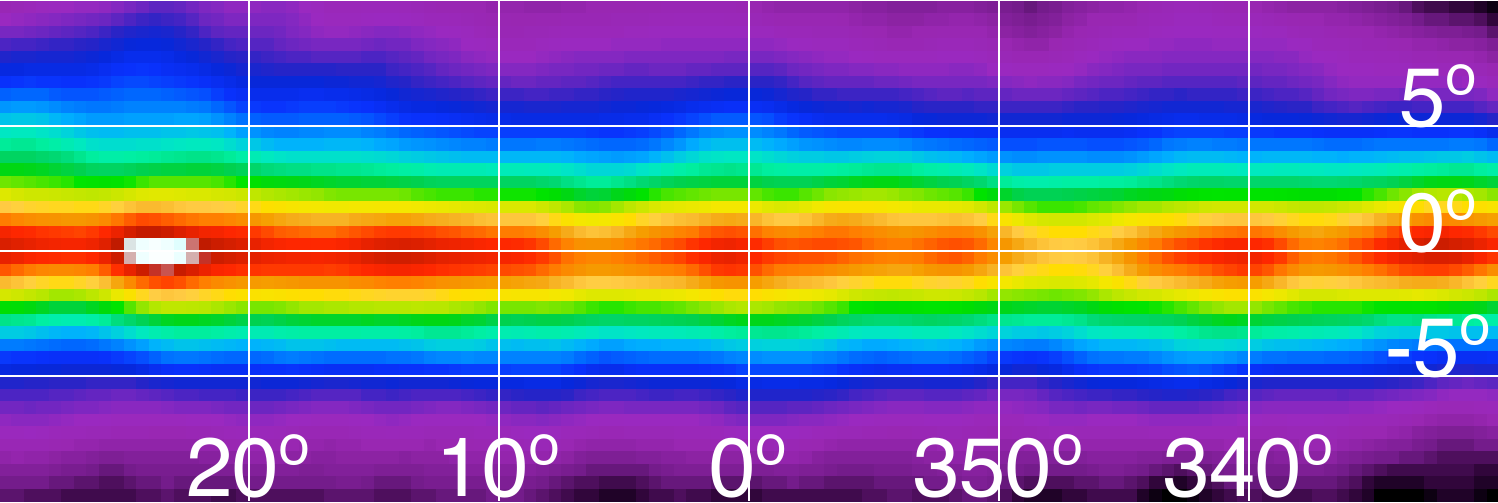} \vspace*{-2.4cm}\\
\includegraphics[height=5.5cm]{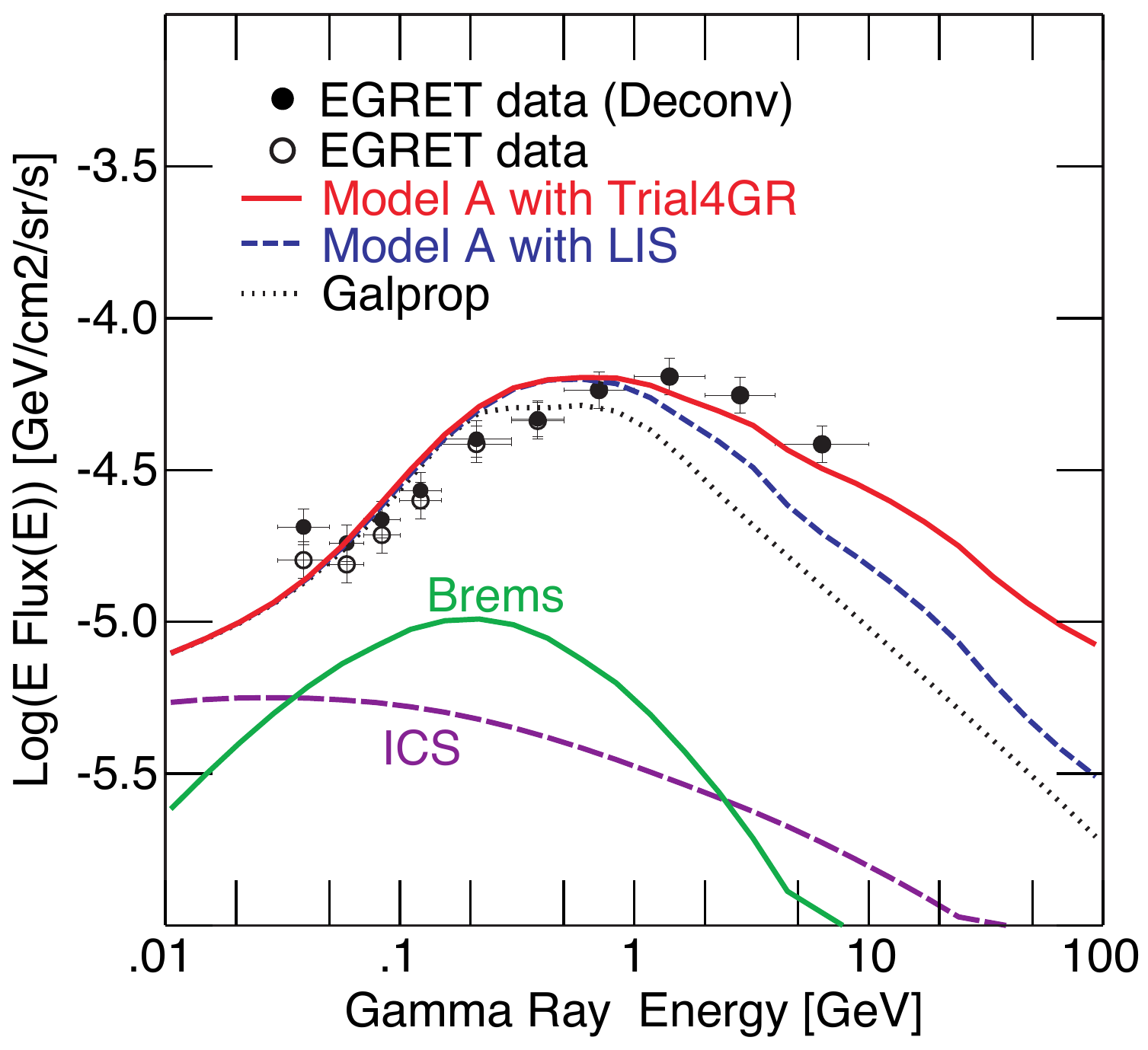} & &
\includegraphics[height=2.4cm]{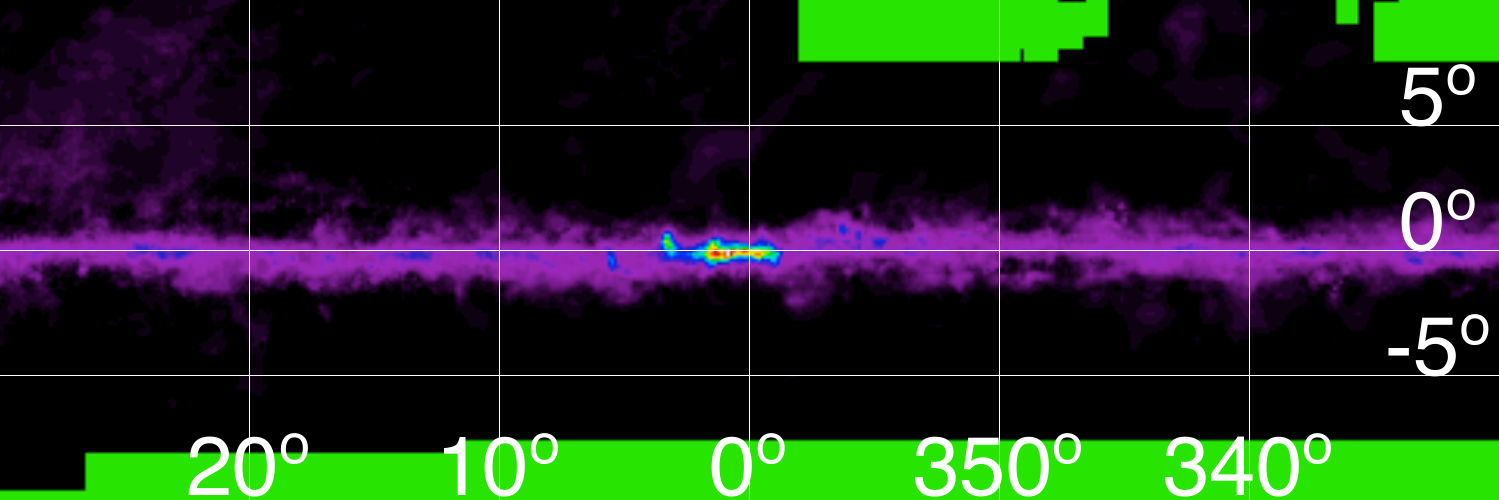}
\end{tabular}
\caption[]
{(a) The diffuse gamma-ray differential spectrum, multiplied by $E^2$ for $330^\circ<\ell<30^\circ$, $|b|<6^\circ$\cite{Kamae05}. Data points are from EGRET observations. Model calculations are superimposed by curves.
(b) EGRET diffuse gamma-ray image in the Galactic center region (top panel) and CO survey map in the same region\cite{Dame01} (bottom pannel). }
\label{fig:Galactic-diffuse}
\end{figure}
Mori\cite{Mori97} fitted the observed spectrum with a simulation code (Pythia) developed for particel physics experiments. 
He noted that this excess can be explained by a harder cosmic-ray spectrum, $1.48\times E^{-2.45}$~cm${}^{-2}$s${}^{-1}$sr${}^{-1}$GeV${}^{-1}$.
He suggested that the cosmic-ray spectrum in the Galactic center region could be harder than that in the solar neighborhood, which was used for the above model calculations.
Kamae {\it et al}\cite{Kamae05} corrected the old nuclear interaction models and demonstrated that 20--50\% more pions are produced compared with old models used by Hunter {\it et al} by correctly accounting for the diffraction dissociation, scaling violation and rising inelastic cross-section in the nuclear interactions as shown in ``Model A with LIS (Local Interstellar Spectrum)'' of Fig.\ref{fig:Galactic-diffuse} (a).
However, the remaining excess still requires a harder cosmic-ray spectrum as shown in ``Model A with Trial4GR'' of Fig.\ref{fig:Galactic-diffuse} (a) where the power law index is set to $-2.5$ instead of $-2.7$ above 20~GeV.

Obviously, better understanding of the cosmic-ray spectrum in the Galactic center region is crucial to solve this problem.
In addition, contamination by unresolved gamma-ray point sources may complicate this problem.
Systematic studies of high-density regions well surveyed by CO observations (both inside and outside of the Galactic plane) and associated gamma-ray emissions are extremely useful. For some cases the CO observation can provide 3D information of the ISM, which can be used in modeling acceleration and propagation of protons and electrons at selected supernova remnants.
Fig.~\ref{fig:Galactic-diffuse}~(b) shows the EGRET diffuse gamma-ray image in the Galactic center region (top panel) and the CO map in the same region\cite{Dame01} (bottom pannel).
For dense molecular clouds, CS survey\cite{Tsuboi99} and other CO transitions can supplement the CO data. Finer and more complete surveys (eg. Ref.~\citenum{Onishi01}) are becoming available in this region. 

\section{Cosmic Rays from Extra-Galactic Sources}
\begin{figure}[hbt]
\centering
\begin{tabular}{ll}
(a) & (b) \\
\includegraphics[height=5.4cm]{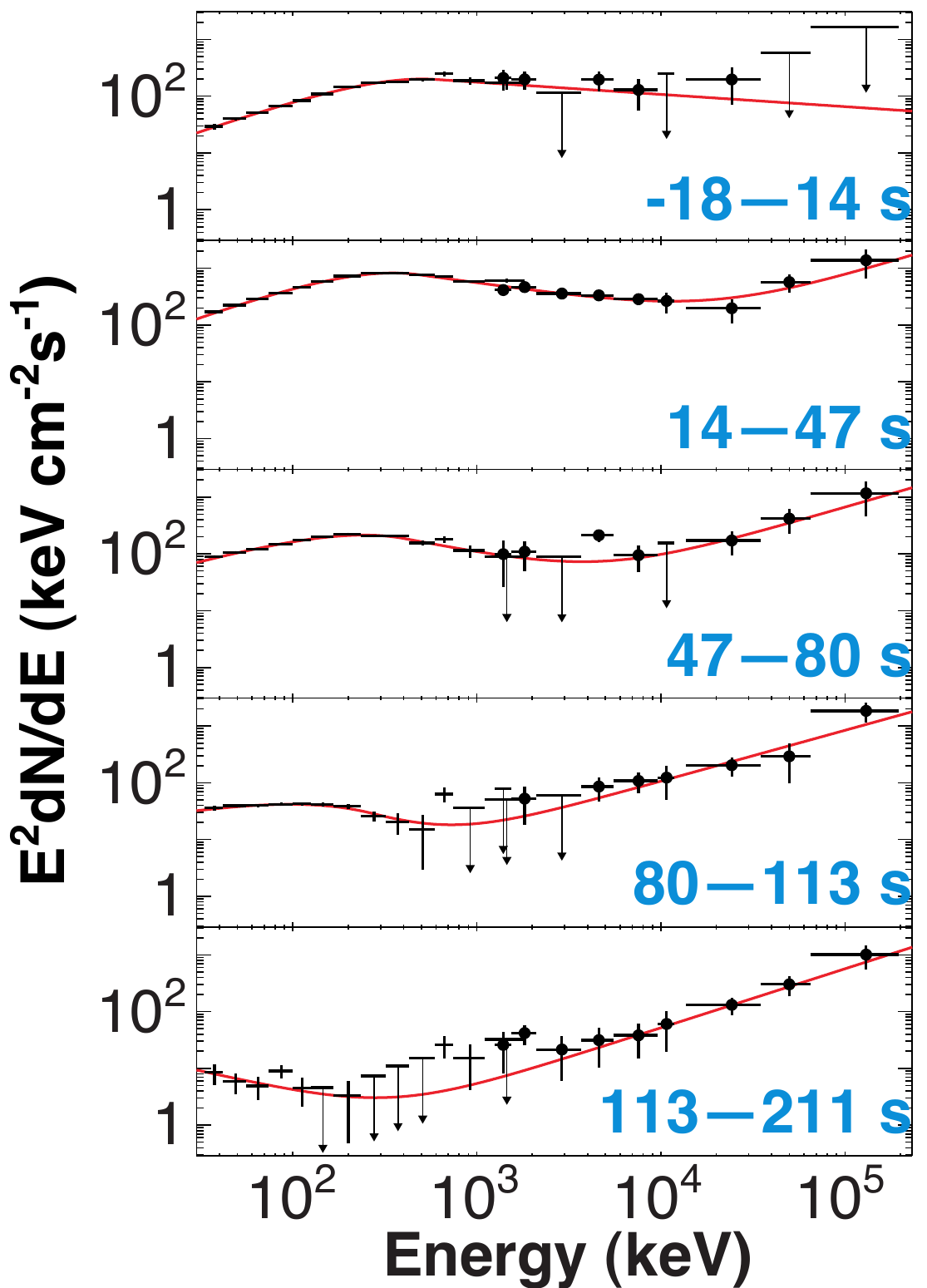}
&\includegraphics[height=5.4cm]{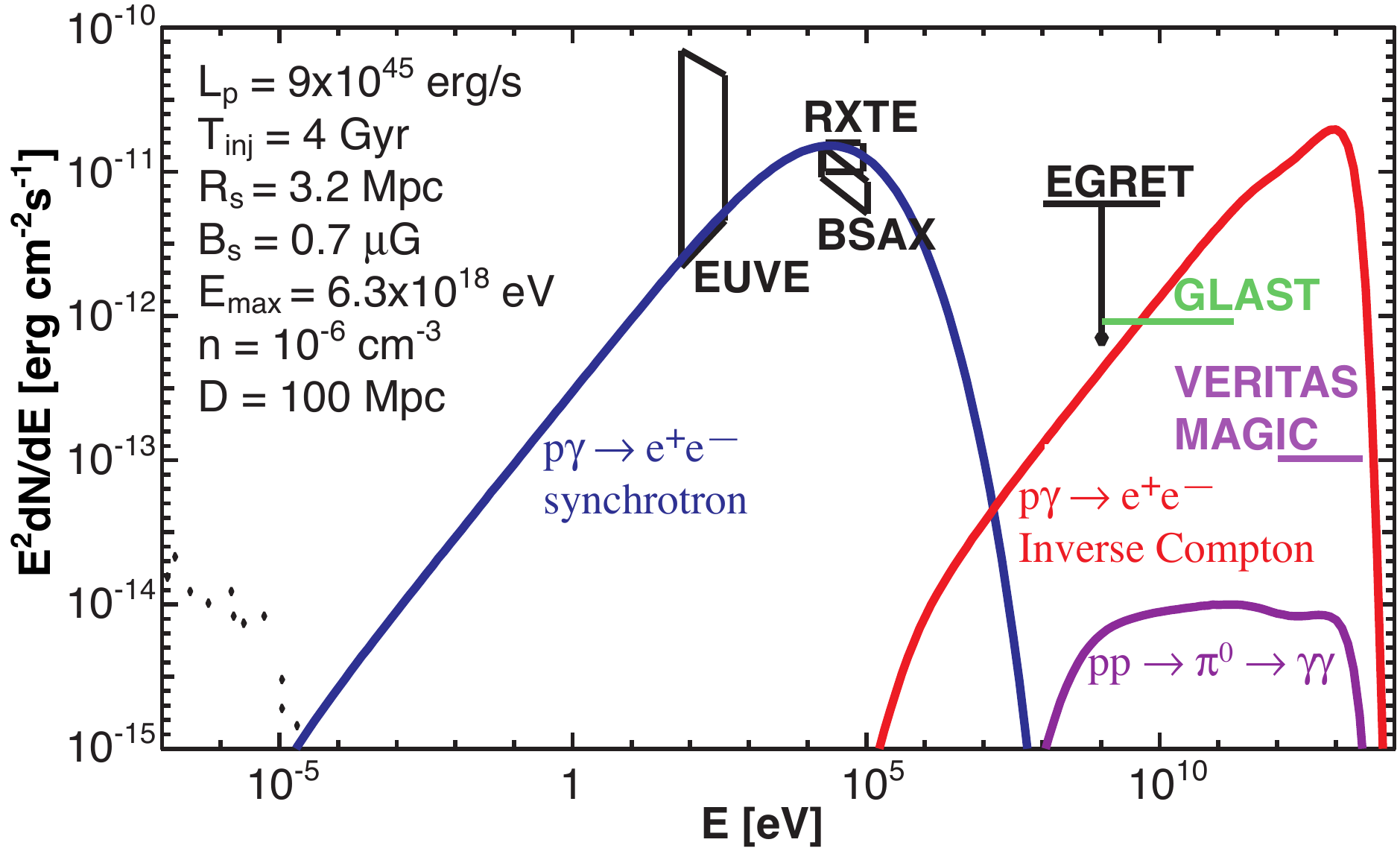}
\end{tabular}
\caption[]
{(a) Time dependent energy fluxes for GRB941017\cite{Gonzalez}. Crosses and circles correspond to BATSE-LAD and EGRET-TASC data respectively.
(b) Model prediction of gamma-ray spectrum for a merging galaxy cluster, Coma\cite{Inoue}. Expected sensitivity limit of GLAST-LAT and IACTs are also displayed.}
\label{fig:extragalactic}
\end{figure}

Gamma-ray bursts (GRBs) are among the most energetic explosions in the Universe, originating possibly in core-collapse of massive stars or binary mergers of neutron stars and/or black-holes. There are theoretical speculations that cosmic rays can be accelerated beyond the ``knee" in GRBs.
We expect that the prompt gamma-ray emissions come from compact regions because of the observed variability (down to milliseconds).
The gamma-ray power spectrum of the GRBs is believed to follow broken power law with the peak energies at around 0.1--1~MeV. This spectral feature is interpreted as due to synchrotron radiation or Compton up-scattering in highly relativistic internal shock regions. 
However, the delayed high energy gamma-ray emission ($E>10$~MeV) observed in the GRB941017\cite{Gonzalez} as shown in Fig.~\ref{fig:extragalactic}~(a) presents a serious challenge to the standard interpretation.
Soft gamma-ray emission decayed in 100~s while the high energy component remained stable. 
Such difference in time structure is difficult to explain with simple electron models.
The photo-hadronic ($p\gamma$) interactions may produce GeV--TeV gamma-ray emissions due to secondary electrons from hadronic and electromagnetic cascades\cite{Razzaque}.
GLAST-LAT will observe many more GRBs like GRB941017 and clarify the origin of high-energy gamma rays. 
Such data will shed new light into understanding the GRB mechanism.
In particular, finding out the peak energy and the power of the high energy component will provide strong contraints on GRB gamma-ray emission models.

Merging galaxy clusters are also considered to be one of candidates for the origin of UHECRs because the Lamor radius can reach Mpc.
UHECRs (protons) accelerated in such acceleration sites will interact with CMB to producing UHE electrons and positrons over cosmological time scale. 
These secondary leptons can produce hard X-rays via synchrotron radiation and gamma rays via Compton up-scattering of CMB photons.
Multi-wavelength observation of merging galaxy clusters in X-ray and gamma-ray bands can constrain such models as demonstrated by Fig.~\ref{fig:extragalactic}~(b)\cite{Inoue}.
It shows a model prediction of X-ray and gamma-ray spectra with upper limits given by existing X-ray observations for the Coma galaxy cluster.
Observation of the Coma with GLAST-LAT (and Imaging Atomospheric \v{C}erenkov Telescopes in the northern hemisphere such as MAGIC or VERITAS) will either confirm or put a stringent constrains to this scenario.

\section{Extended Source Analysis for GLAST}
The sources discussed above are most likely spatially extended.
Modest PSF (point-spread function) and photon statistics of the GLAST-LAT compared with X-ray instruments present challenge to studies on these extended sources.
Morphological feature of TeV sources has played a critical role in identifying all SNRs and PWNe (Pulsar Wind Nebulae) in H.E.S.S. analyses.
Several TeV sources remain unidentified probably due to limited spatial resolution in the TeV band.
The PSF of GLAST-LAT is often comparable to size of extended sources and morphology of sources may not be evident.
In addition, the presence of the Galactic diffuse gamma-ray background makes it even more difficult to resolve extended sources.
The fact that roughly half of 271 gamma-ray sources\cite{Hartman99} detected by EGRET remain unidentified underscores the challenge GLAST-LAT will face.
We note that many of the unidentified EGRET sources are located at low Galactic latitude and are considered to be of Galactic origin.
Furthermore, extended sources will be contaminated with contribution from nearby sources.

The iterative image restoration technique originally suggested by Richardson and Lucy\cite{Richardson72,Lucy74} has potential to improve the quality of GLAST-LAT and EGRET images.
Hook and Lucy also developed the dual channel method\cite{Hook94} to account for the effect of point sources in the restoration process, which is effective in removing point-source contibution from a diffuse image.
However, this technique tends to suffer from amplification of the Poisson noise as the number of iteration is increased.
Efforts have been made to define objective criteria to stop the iteration before such effects become prominent\cite{Lucy94}.
Charalabides {\it et al}.\cite{Charalabides} applied this technique to the EGRET images using a wavelet filtering to suppress the effect of Poisson noise\cite{Donoho93}.
However, the results suffered from rapidly-varying and energy-dependent PSF.
We have developed a novel image restoration technique that incorporates PSF calculated for each event to address this problem.
This technique will be critical for the analysis of extended sources in the GLAST-LAT since its PSF varies by two orders of magnitude over the wide energy band, 0.1--300~GeV.

\subsection{Image Restoration Technique}
The observed image $\tilde{\phi}(x)$ is a convolution of the true image $\psi(\xi)$ and the instrument response function $P(x|\xi)$, where $P(x|\xi)$ is the probability that the photon is observed at $x$ when the true position is $\xi$.
Based on a Bayesian approach, the original image $\psi(\xi)$ can be obtained iteratively:
\begin{equation}
\psi^{r+1}(\xi) = \int\tilde{\phi}(x)\frac{\psi^r(\xi)P(x|\xi)}{\int P(x|\zeta)\psi^r(\zeta)d\zeta}dx.
\end{equation}
It is proven mathematically that $\psi^{r+1}(\xi)$ yields larger likelihood than $\psi^{r}(\xi)$ at each iteration.
We use wavelet denoising to suppress the noise in the residual between the observed and expected image in each iteration\cite{Starck94}.
The residual, $\rho^r(x)$, is defined as $\rho^r(x) \equiv \tilde{\phi}(x) - \phi^r(x)$, where $\phi^r(x)$ is the convolved image:
\begin{equation}
\phi^r(x) \equiv \int P(x|\zeta)\psi^r(\zeta)d\zeta.
\end{equation}
The residual is decomposed by the ``\`a trous'' wavelet algorithm into $J$ components:
\begin{equation}
\rho^r(x) = c_J(x) + \sum_{j=1}^{J}w_j(x),
\end{equation}
where $c_J(x)$ is the last smoothed image and $w_j$ denotes the wavelet component at scale $j$.
Denoising is performed by requiring the values to be above noise in each component: the filtered residual is written as
\begin{equation}
\overline{\rho}^r(x) = c_J(x) + \sum_{j=1}^{J}M_j(x)w_j(x)\ ,
\end{equation}
where $M_j(x)$ is 0 if $w_j(x)$ is consistent with noise and 1 otherwise (hard-thresholding).
We currently use the 99\% Poisson probability as the consistency threshold.
The filtered residual is incorporated in the deconvolution process by replacing 
the observed image $\tilde{\phi}(x) \equiv {\rho}^r(x) + \phi^r(x)$ with the filtered image $\overline{\phi}^r(x) \equiv \overline{\rho}^r(x) + \phi^r(x)$,
\begin{equation}
\psi^{r+1}(\xi) = \int\overline{\phi}^r(x)\frac{\psi^r(\xi)P(x|\xi)}{\int P(x|\zeta)\psi^r(\zeta)d\zeta}dx. 
\end{equation}

When we have known point sources, the probability density function, $\psi^{r}(\xi)$, can be decomposed into two components: $\psi^r(\xi) = \psi_{\mathrm PS}^r(\xi) + \psi_{\mathrm DF}^r(\xi)$, where $\psi_{\mathrm PS}^r(\xi)$ represents a point source component (mostly zero apart from $\delta$-functions at the positions of the point sources) while $\psi_{\mathrm DF}^r(\xi)$ represents a diffuse component.
Wavelet filtering is applied only to the diffuse component.
This dual channel method enables us to incorporate sharp point sources while keeping the smooth diffuse component.
This effectively works as point source subtraction without negative values in $\psi_{\mathrm DF}^{r}(\xi)$ (straight point source subtraction often causes negative values due to noise).

This image restoration technique can be generalized for the case where the PSF is assigned for each event as:
\begin{equation}
\psi^{r+1}(\xi) = \frac{1}{N}\psi^r(\xi)\sum_{k=1}^{N}\frac{P_k(x_k|\xi)}{\int P_k(x_k|\zeta)\psi^r(\zeta)d\zeta},
\end{equation}
where $x_k$ is the observed position of the $k$th event and $P_k(x_k|\xi)$ is the probability to observe the event at $x_k$ when the true position is $\xi$.
In this case, the convolved image is defined as
\begin{equation}
\phi^r(x) = \int \Pi(x|\zeta)\psi^r(\zeta)d\zeta,
\end{equation}
where $\Pi(x|\zeta)$ is the weighted average of the instrument response at a given position $x$ and defined as:
\begin{equation}
\Pi(x|\zeta) = \frac{\sum_{k=1}^{N}P_k(x|\zeta)P_k(x_k|x)}{\sum_{k=1}^{N} P_k(x_k|x)}.
\end{equation}
The resulting residual, $\rho^r(x)$, can be filtered with the same wavelet denoising method as above to obtain $\overline{\rho}^r(x)$.
The filtered image $\overline{\phi}^r(x) \equiv \overline{\rho}^r(x) + \phi^r(x)$ can be incorporated in the iteration process as:
\begin{equation}
\psi^{r+1}(\xi) = \frac{1}{N}\psi^r(\xi)\sum_{k=1}^{N}\frac{P_k(x_k|\xi)\overline{\phi}^r(x_k)/\tilde{\phi}(x_k)}{\int P_k(x_k|\zeta)\psi^r(\zeta)d\zeta},
\end{equation}
These generalized formula described here are completely equivalent to the non-generalized version if $P_k(x_k|\xi)$ is common for all events.

\subsection{Application to EGRET Image Analysis}
In this section, we demonstrate our image restoration technique by analyzing three fields observed by EGRET.
The PSF is calculated for each event as a convolution of PSFs in different energy bins, according to the energy dispersion of the given energy measurement.
Only photons with energy greater than 0.5~GeV are used in this analysis in order to minimize the processing time without sacrificing the image quality.
(PSF in this energy region varies from $\sim2^\circ$ to $0.5^\circ$.)
Lower-energy photons do not contribute to improving the quality of the restored image.
It does not preclude the use of lower-energy photons in the future analysis, in particular for spectral analysis, since it does not degrade the image quality either since such photons are weighted appropriately via a large PSF.
The results presented in this paper are still preliminary and should not be used for any quantitative analysis.

\subsubsection{Large Magellanic Cloud}
\begin{figure}[hb]
\centering
  \includegraphics[height=4.7cm]{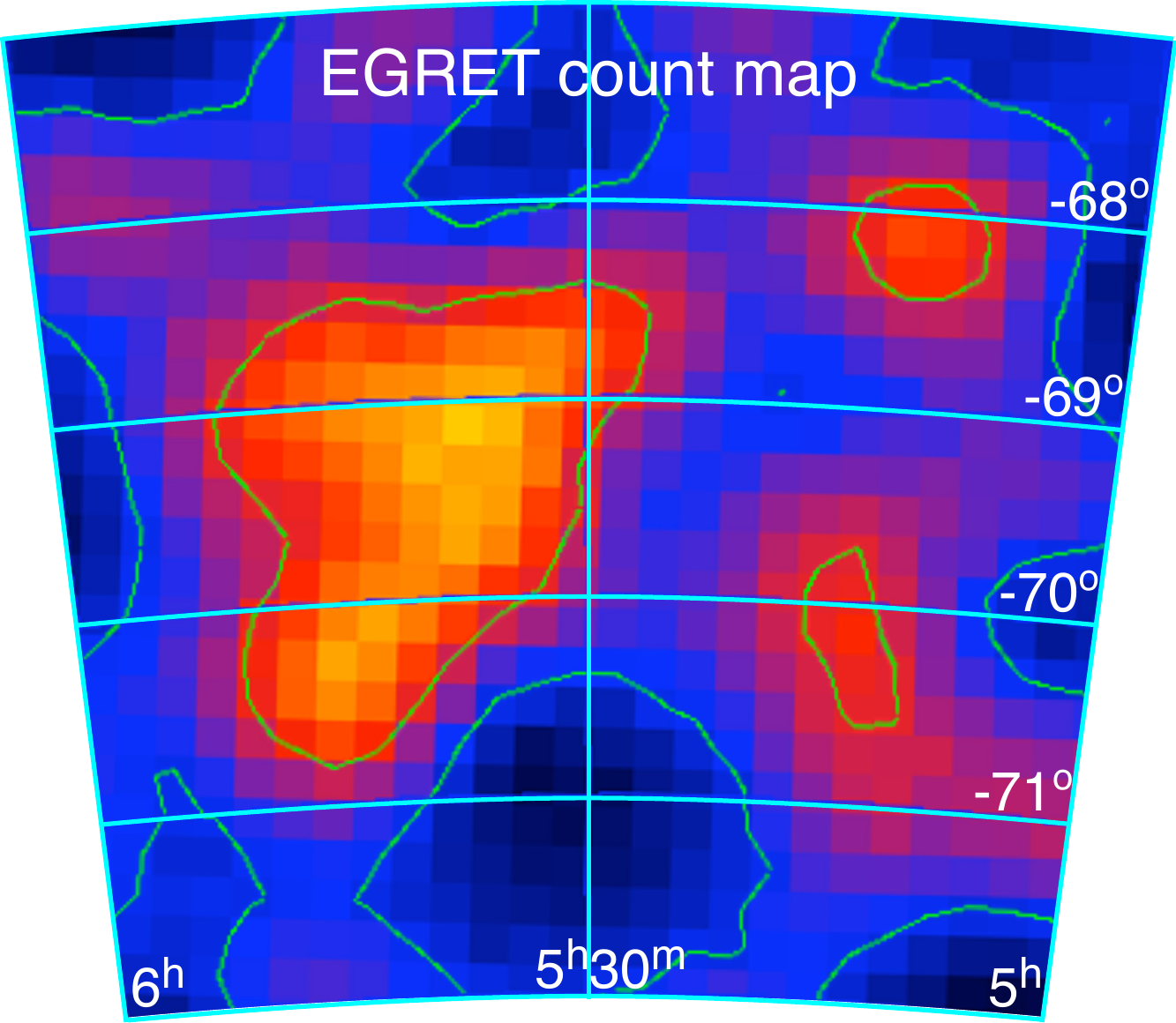} 
  \includegraphics[height=4.7cm]{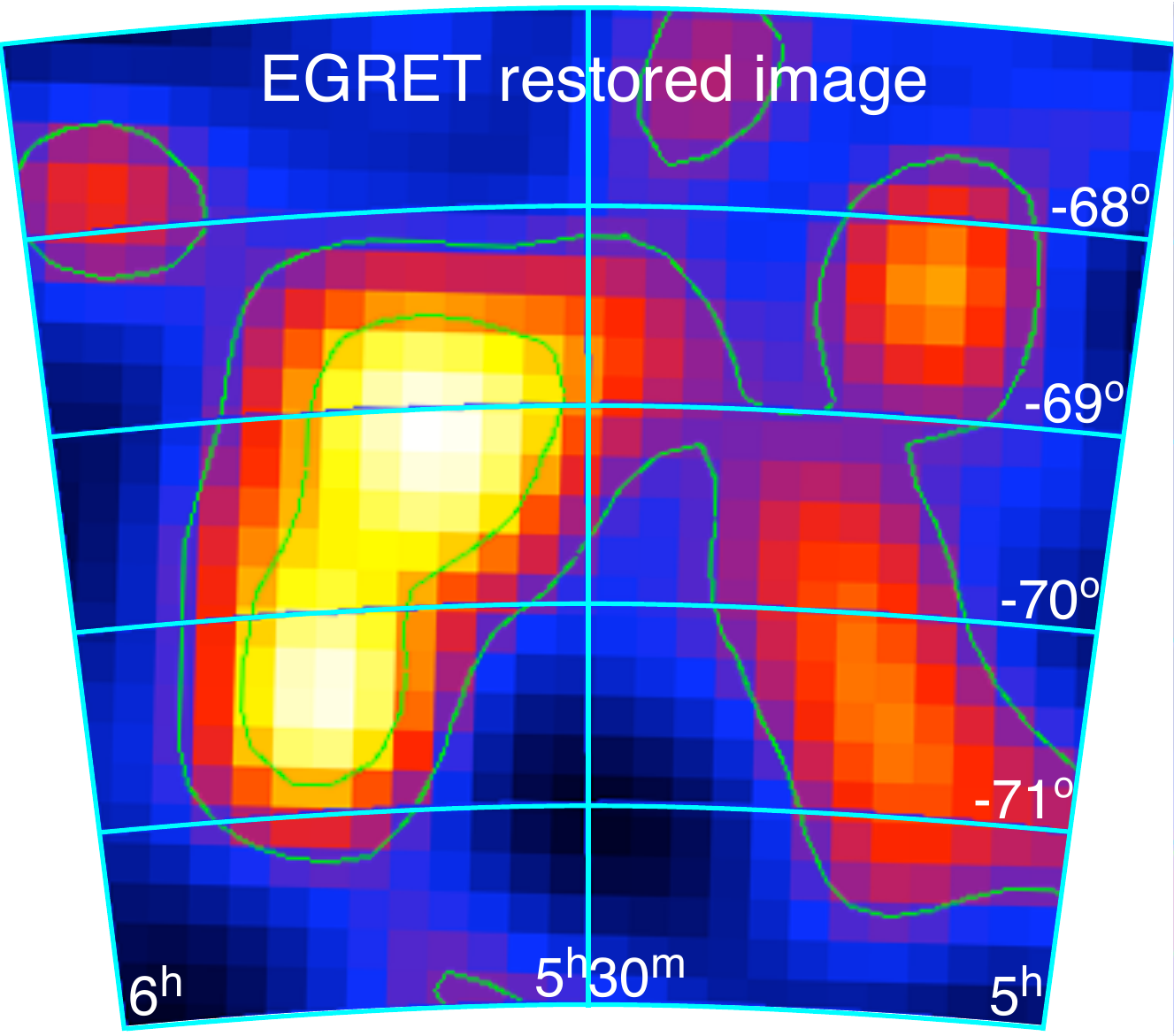}
  \includegraphics[height=4.7cm]{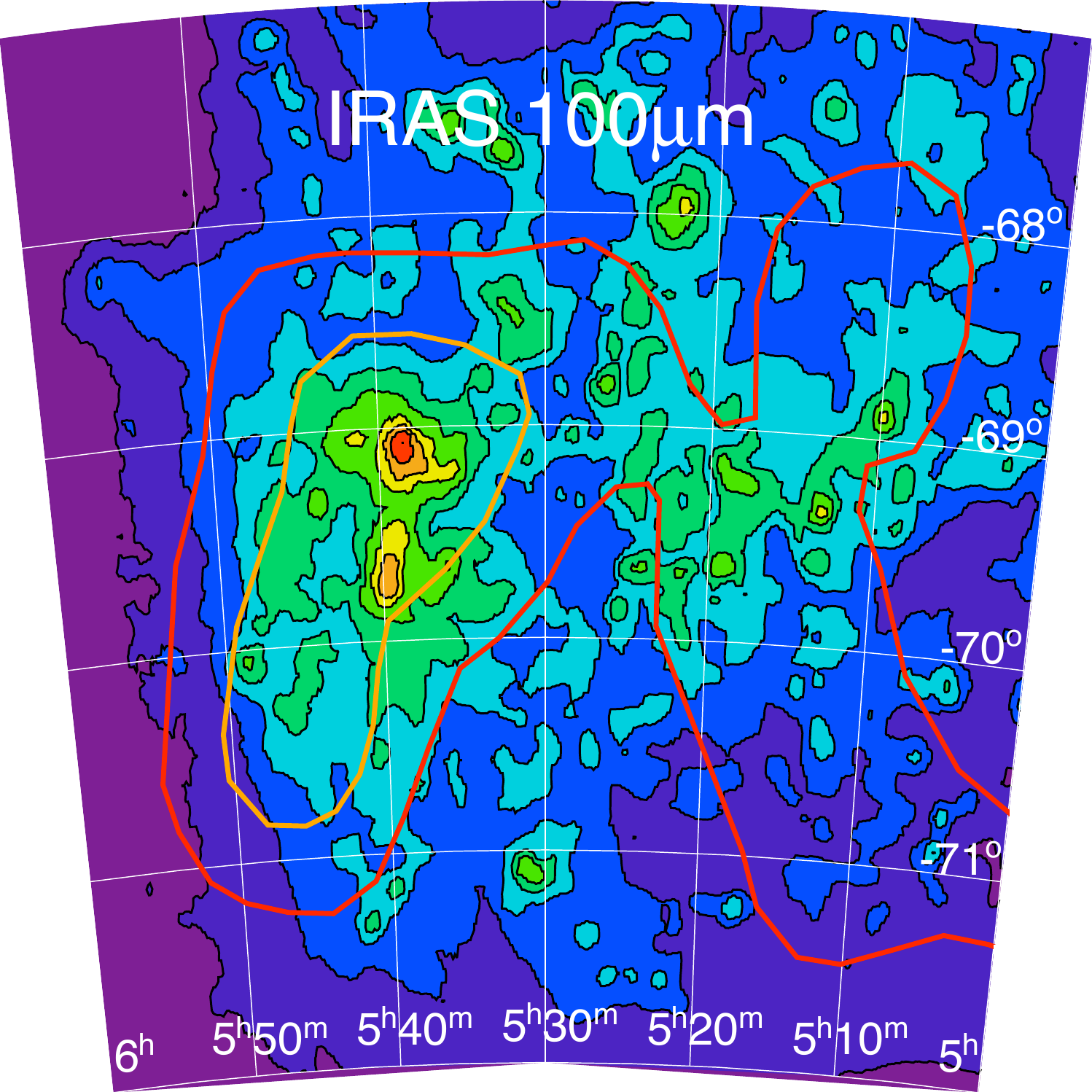}
  \caption{\label{fig:LMC}Smoothed count map (left), deconvolved image (middle) of the LMC by EGRET, and the IRAS LMC image overlaid with contour of the deconvolved EGRET image (right).}
\end{figure}

The LMC (Large Magellanic Cloud) is one of few extended sources resolved by EGRET. 
Fig.~\ref{fig:LMC} shows a smoothed photon count map of the LMC in the left panel, a restored image of the LMC in the middle panel, and the IRAS LMC image overlaid with the contour of the restored EGRET image in the right panel.
The restored EGRET image is apparently sharper with better contrast than the original count map while avoiding enhancement of spurious structures (sometimes suppressing them) in the background (The color scale is common for both images).
The bright spot of the restored image is well correlated with a dense region indicated by the IRAS image, which traces dusty interstellar clouds and associated star-forming regions.

\subsubsection{3EG J1234-1318}
\begin{figure}[thb]
\centering
  \includegraphics[height=4.9cm]{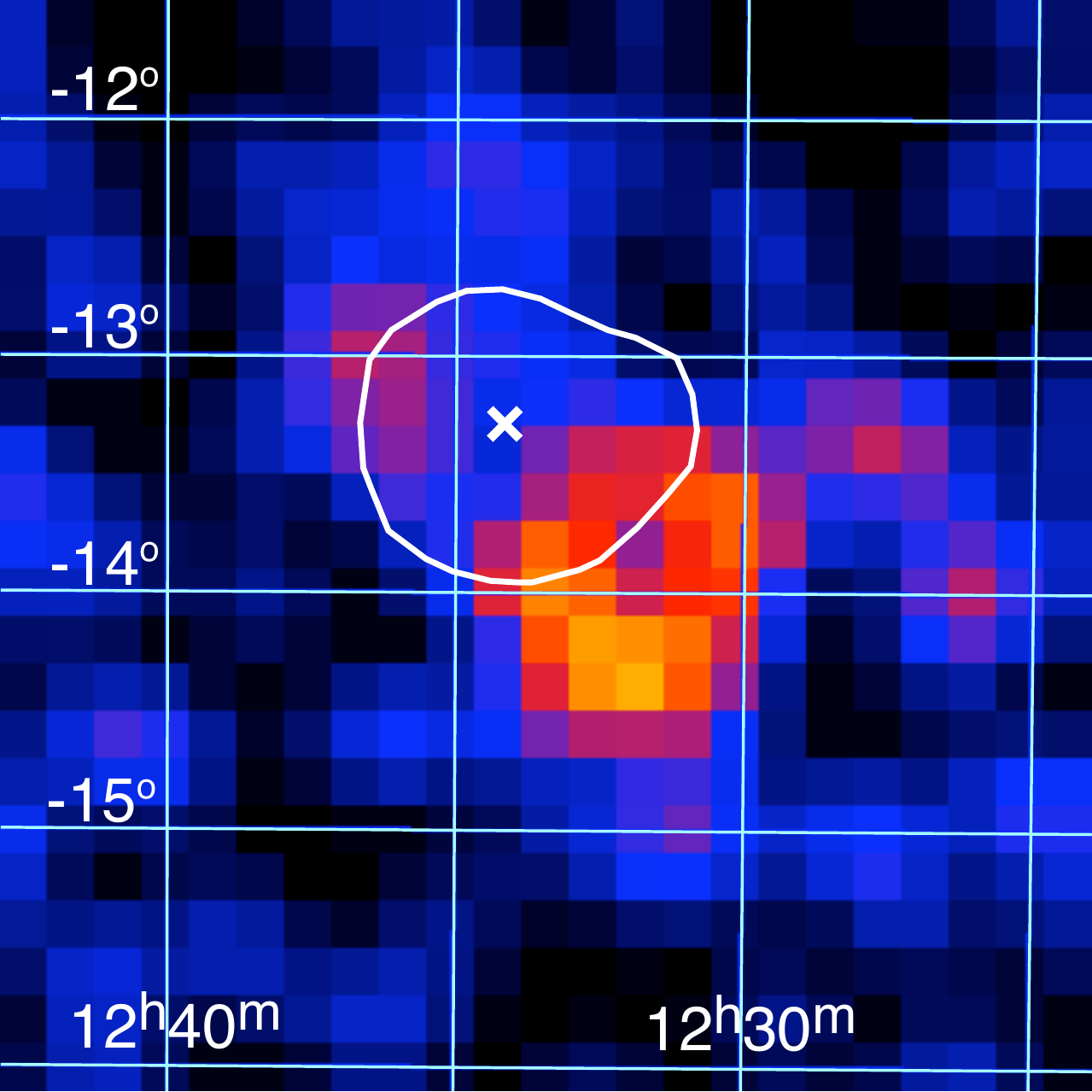}
  \includegraphics[height=4.9cm]{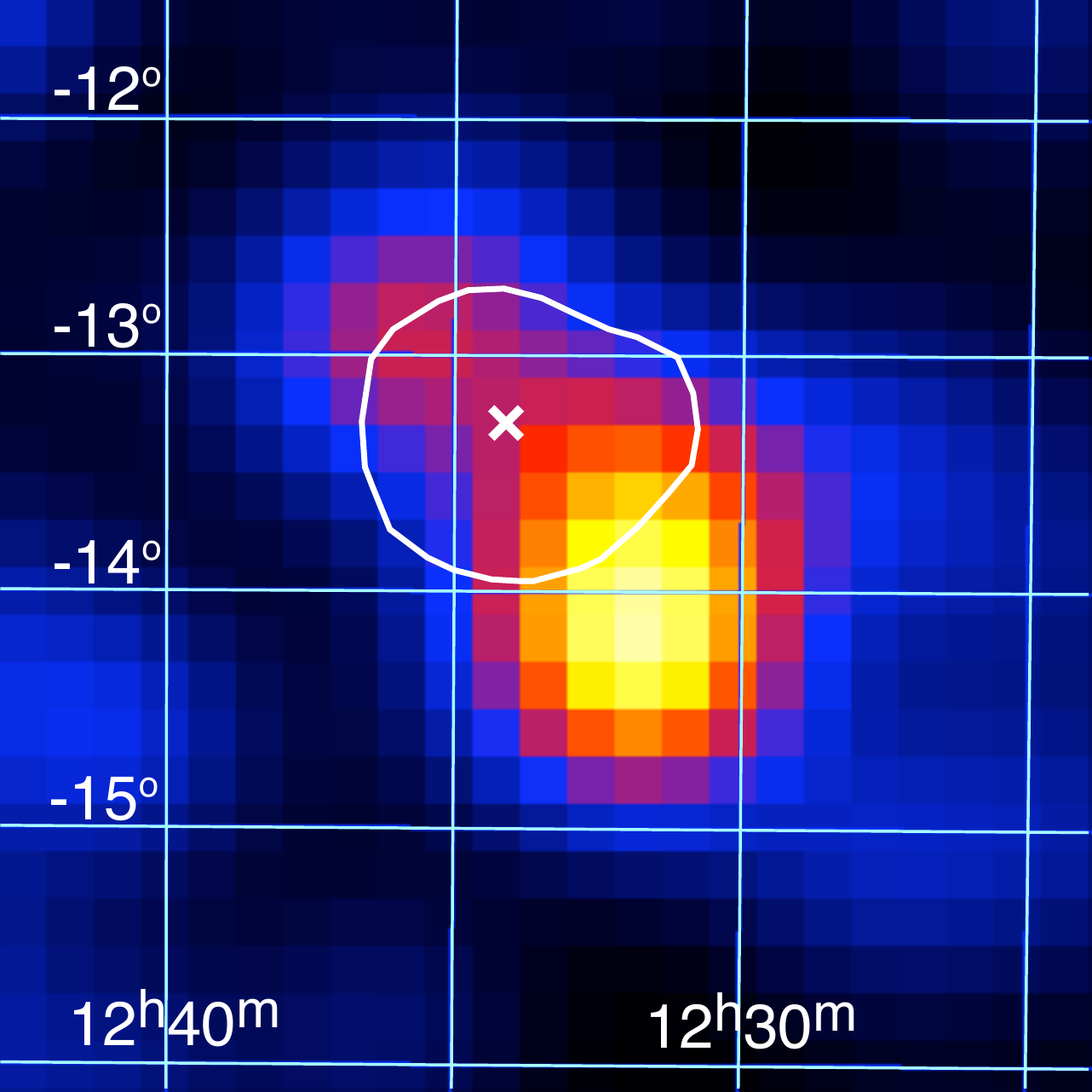}
  \caption{\label{fig:EG1234-1318}Smoothed count map (left) and deconvolved image (right) of the source 3EG J1234-1318 by EGRET.}
\end{figure}

We also analyzed several unidentified EGRET sources at high Galactic latitude, where the Galactic diffuse background is very low.
We found one source that is inconsistent with one point source as shown in Fig.~\ref{fig:EG1234-1318}
The left panel shows the smoothed count map of the 3EG J1234-1318 while the right panel shows the restored image.
A white cross and a contour in each image indicate the source location and its 90\% confidence level contour in the original 3EG catalog (The 3EG catalog is produced using photons with $E>0.1$~GeV).
The restored image has much better contrast and clearly shows an elongated structure 
indicating the presence of either an extended source, or two point sources separated by $\sim$1.5${}^{\circ}$.
Since the Galactic latitude of this EGRET signal is $\sim$50${}^{\circ}$\, it is very likely to be two extra-Galactic point sources.
If such is the case, the preliminary location of a {\em new} EGRET point source would be 
(RA, DEC)=($187.9$${}^{\circ}$, $-14.1$${}^{\circ}$) for the brighter one and ($188.9$${}^{\circ}$, $-13.0$${}^{\circ}$) for the 
fainter one instead of the original position, ($188.5$${}^{\circ}$, $-13.3$${}^{\circ}$).
Another possibility is that this is due to a nearby molecular cloud interacting with cosmic rays, which could be extremely useful to obtain cosmic-ray flux (both spectrum and intensity) in that region.
This result demonstrates that our image restoration technique can be quite effective in identification of extended sources and source confusion.

\subsubsection{Galactic Plane}
\begin{figure}[hb]
\centering
\begin{tabular}{cc}
  \includegraphics[height=4.9cm]{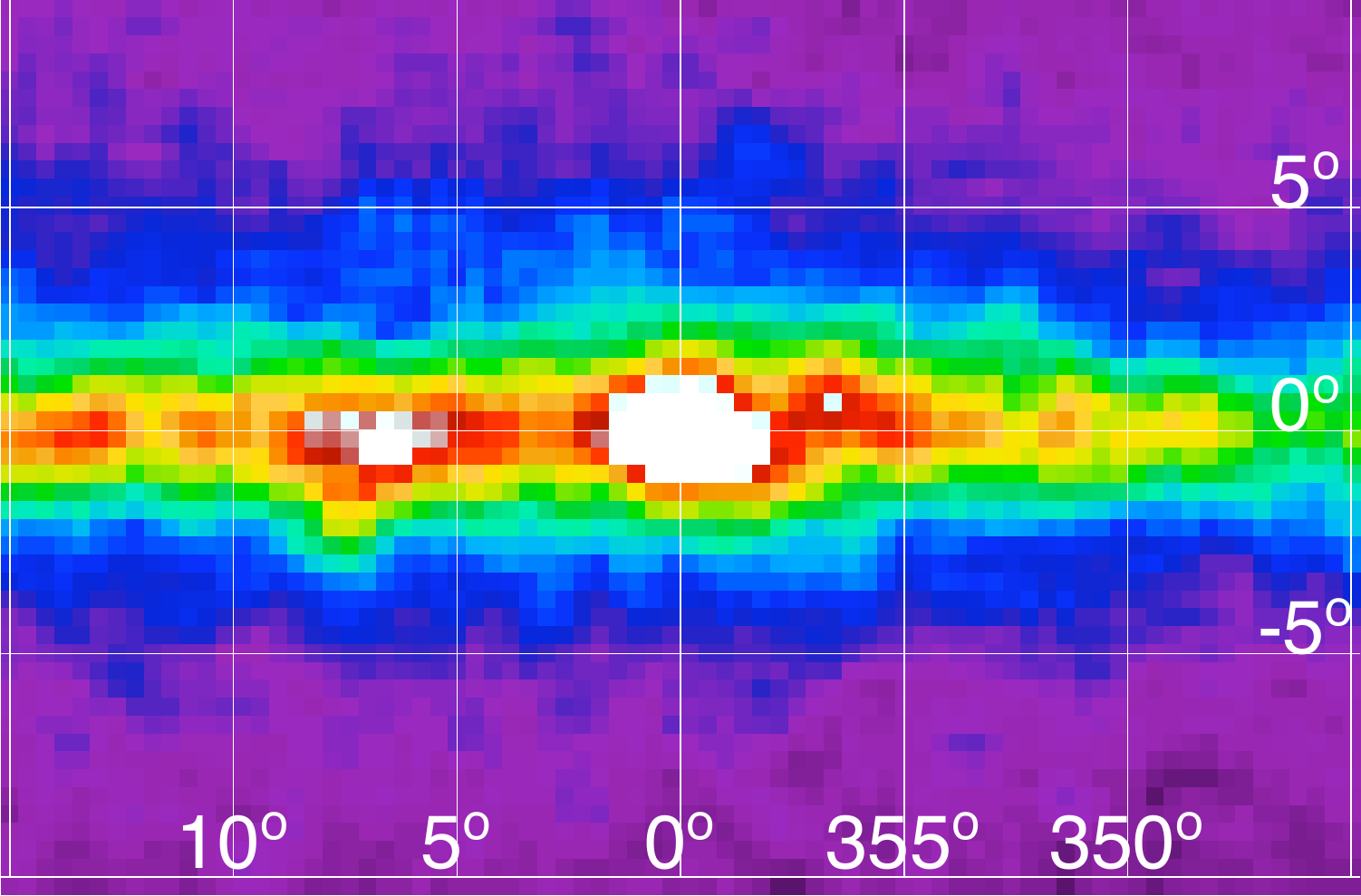} &
  \includegraphics[height=4.9cm]{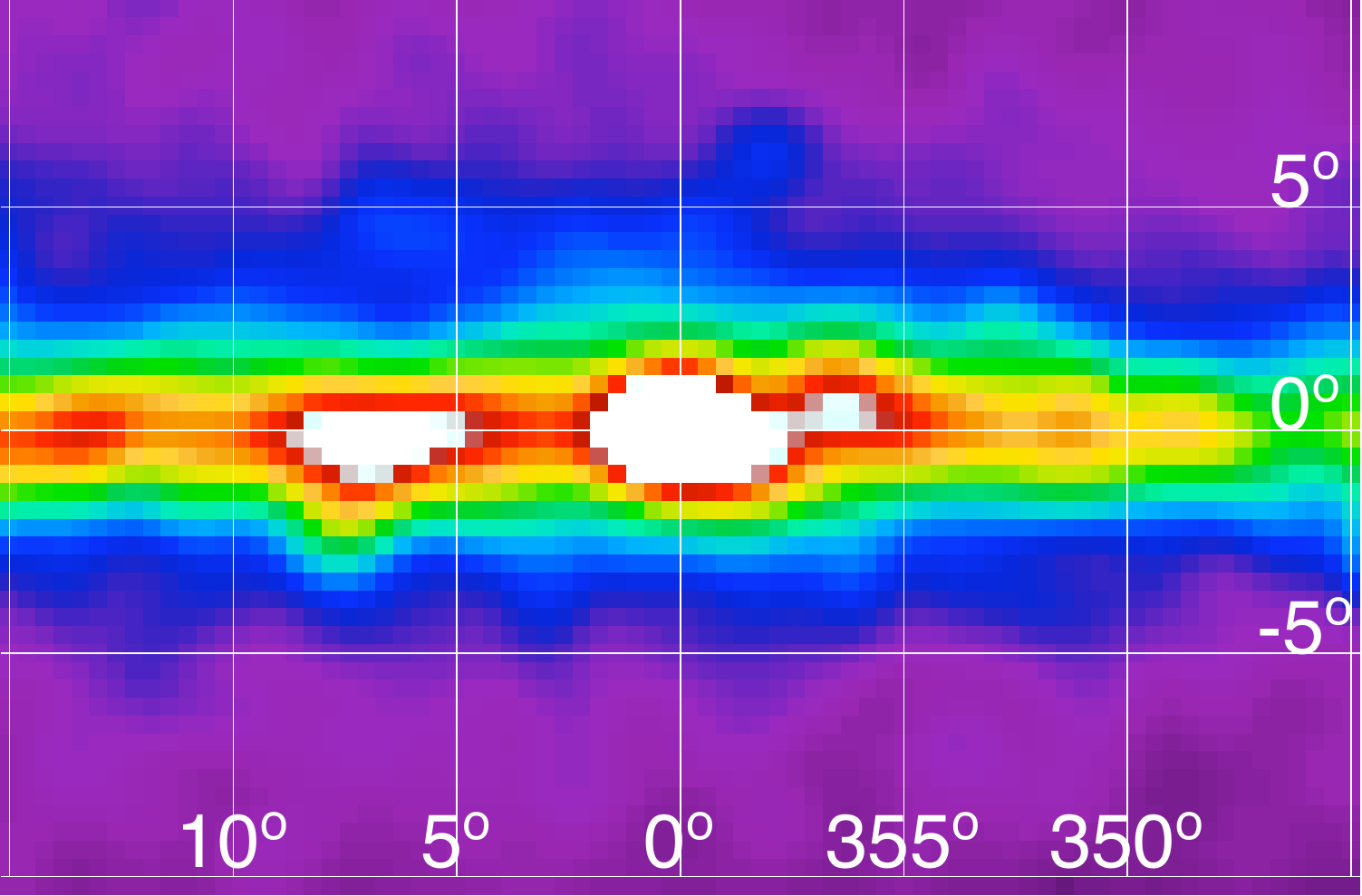} \\
  \includegraphics[height=4.9cm]{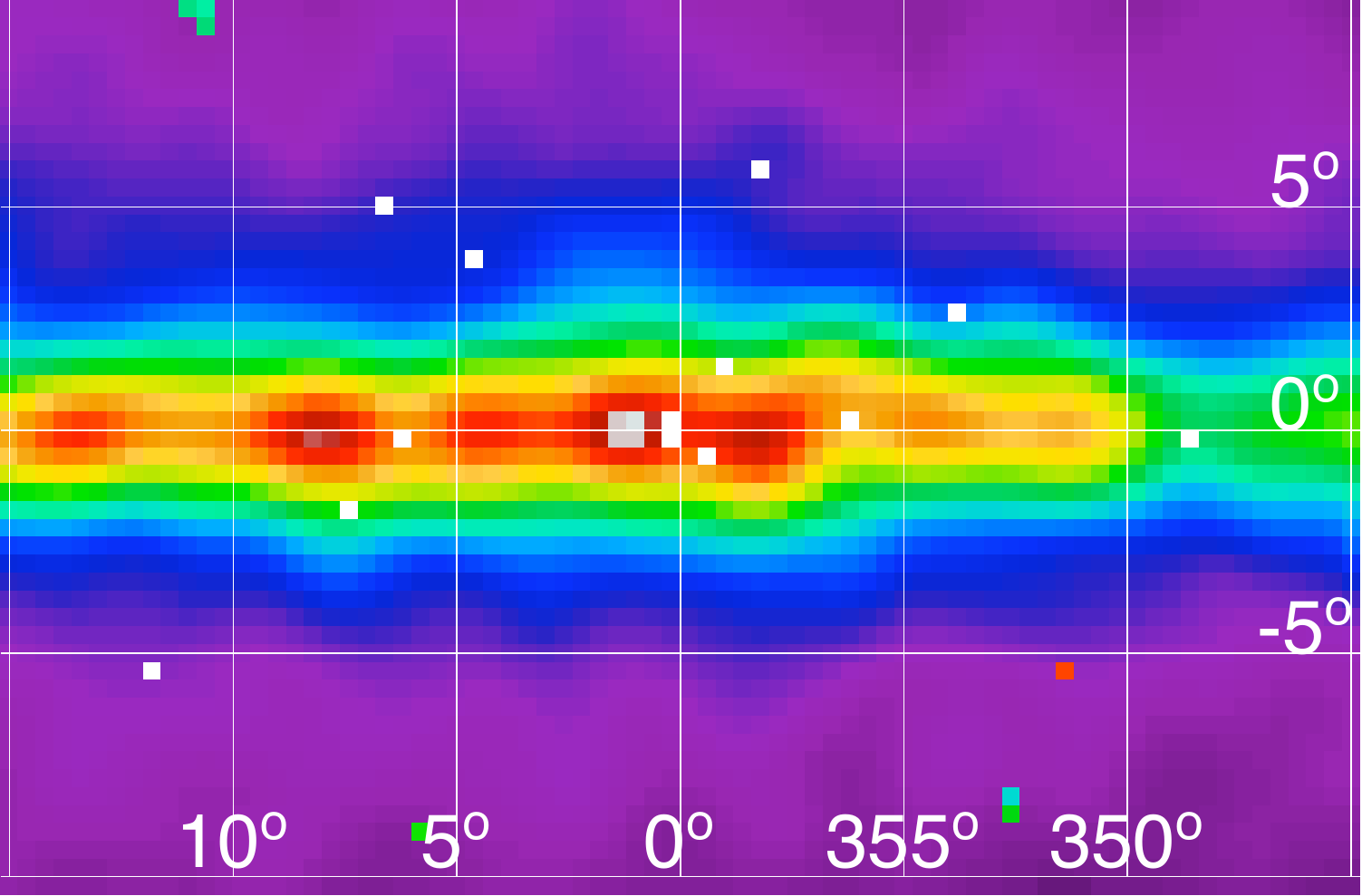} &
  \includegraphics[height=4.9cm]{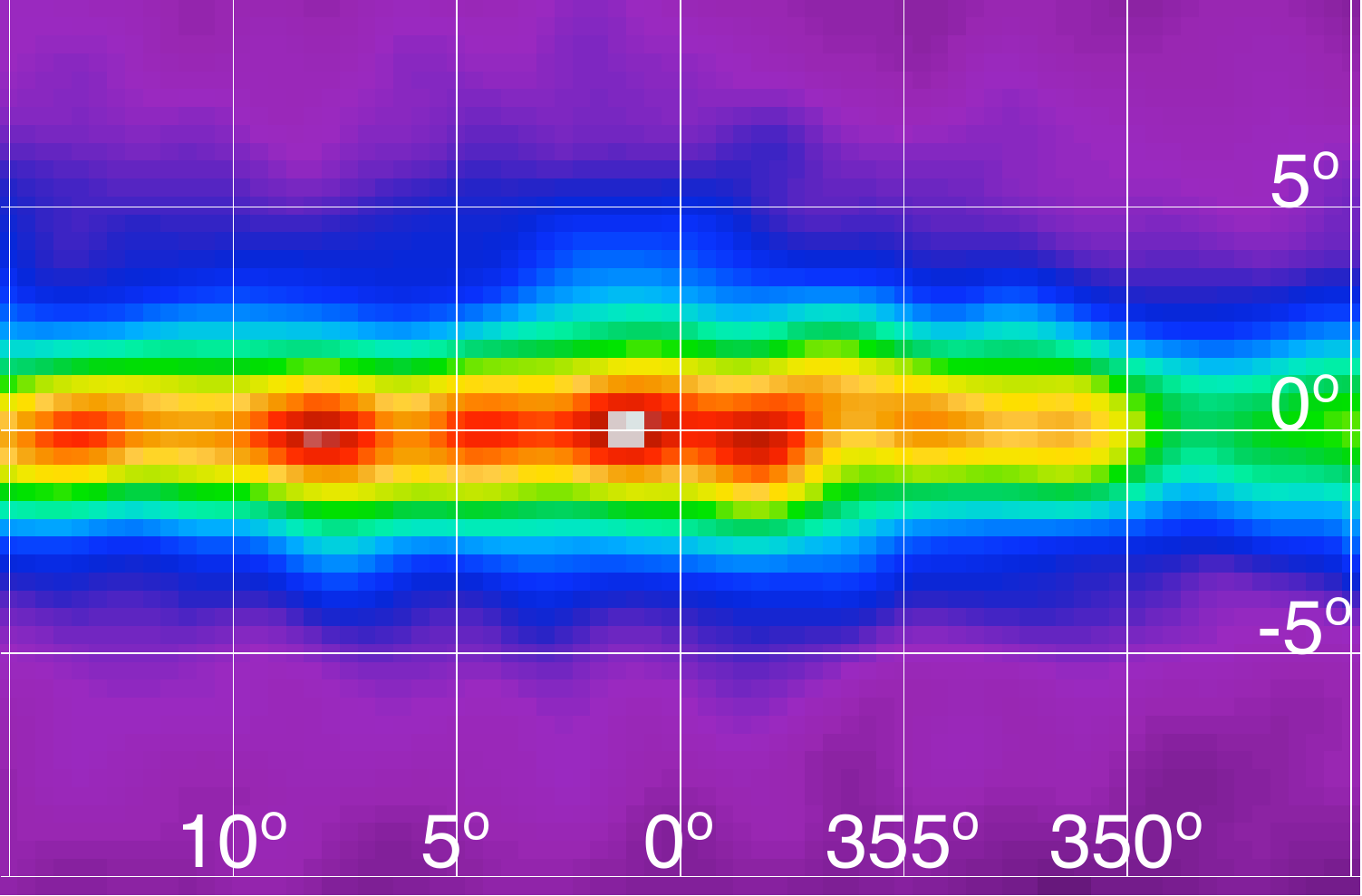}
\end{tabular}
  \caption{\label{fig:Galactic-plane}Smoothed count map (top left), simple deconvolved image (top right), deconvolved image with known point sources included by the dual channel method (bottom left), and  deconvolved image of the only diffuse channel (bottom right) of the Galactic plane around the Galactic center by EGRET.}
\end{figure}

Although the Milky Way can be resolved by the EGRET due to its large size, smaller scale structure in the Galactic plane is not easy to resolve due to the large EGRET PSF and contamination by the point sources.
Our image restoration technique can reduce these effects as shown in Fig.~\ref{fig:Galactic-plane}
The top left panel shows the smoothed count map, the top right panel shows the simple deconvolved image (no dual channel method), the bottom left panel shows the deconvolved image with known point sources included by the dual channel method, and the bottom right panel shows the deconvolved image of the only diffuse channel from the previous image, which is equivalent to the point-source removal.
The final EGRET image of the Galactic plane has smooth background and less structures associated with ``point sources'' that may or may not be true point sources.

\section{Conclusions}
GeV gamma-ray observations by GLAST-LAT will shed light on the origin, propagation and interaction with ISM of cosmic rays.
Specifically, the GLAST-LAT will conclusively distinguish leptonic and hadronic models of gamma-ray emissions in SNRs RX J1713.7-3946 and RX J0852.0-4622.
The GLAST-LAT may give clues on understanding the cosmic-ray acceleration in extra-Galactic sources such as GRBs and clusters of galaxies.

In order to facilitate the analysis of extended sources with limited PSF expected for the GLAST-LAT, generalization of a widely-used image restoration technique is introduced to incorporate the strongly energy-dependent PSF in the EGRET and GLAST-LAT although it is not trivial since the wavelet filtering is performed on pixelated images.
We demonstrated the effectiveness of our technique using three fields observed by EGRET.
Our technique is confirmed to be useful to obtain sharper image of extended sources, to identify source confusion, and to cleanly remove point sources from the image of extended sources.

This work was supported in part by the U.S. Department of Energy under Grant DE-AC02-76SF00515.

\end{document}